\documentclass{article}[10pt]

\usepackage{graphics}

\newcommand{\bea}{\begin{eqnarray}}
\newcommand{\eea}{\end{eqnarray}}
\newcommand{\be}{\begin{equation}}
\newcommand{\ee}{\end{equation}}

\def\be{\begin{eqnarray}}
\def\ee{\end{eqnarray}}
\def\bd{\begin{displaymath}}
\def\ed{\end{displaymath}}
\def\nn{\nonumber}

\def\etal{{\em et al.}}
\def\ADNDT{{\em At. Data. Nucl. Data. Tables }}

\def\RMP{{\em Rev. Mod. Phys. }}
\def\NP{{\em Nucl. Phys. }}
\def\PR{{\em Phys. Rev. }}
\def\PRL{{\em Phys. Rev. Lett. }}
\def\PL{{\em Phys. Lett. }}
\def\jpg{{\em J. Phys. G: Nucl. Part. Phys. }}
\def\EPJ{{\em Eur. Phys. J. }}
\def\ZP{{\em Z. Phys. }}

\begin{document}

\title{A new phenomenological formula for ground state binding energies}

\author{G. Gangopadhyay\\
Department of Physics, University of Calcutta\\
92, Acharya Prafulla Chandra Road, Kolkata-700 009, India\\
email: ggphy@caluniv.ac.in}
\date{}
\maketitle


\begin{abstract}
A phenomenological formula based on liquid drop model has been proposed for 
ground state binding energies of nuclei. The effect due to bunching of single 
particle levels has been incorporated through a term resembling the one-body 
Hamiltonian. The effect of n-p interaction has been included through a function
of valence nucleons. A total of 50 parameters has been used in the present 
calculation. The r.m.s. deviation for the binding energy values for 2140 nuclei comes 
out to be 0.376 MeV, and that for 1091 alpha decay energies is
0.284 MeV. The correspondence with the conventional liquid drop model is 
discussed.
\end{abstract}


\section{Introduction}

The charged liquid drop model of the nucleus, in combination with various 
modifications, has very successfully been applied to describe nuclear ground 
state binding energy and many other ground state properties\cite{rmp}. 
Most of the modern global mass formulas retain the original  
Bethe-Weizs\"{a}cker (BW) form as basis and add certain effects such as shell correction,
deformation, surface symmetry terms, etc. Fully microscopic approaches such as
Skyrme Hartree Fock or Relativistic Mean Field methods also have been 
employed to estimate the mass values through out the periodic table. Local 
formulas, as the name suggests, predict the unknown mass of a nucleus by 
extrapolation from  the known masses of neighbouring nuclei.

Fully microscopic approaches are yet to reach the accuracy achieved by
microscopic-macroscopic mass formula. Skyrme Hartree Fock mass values
show an r.m.s. deviation slightly less than 0.6 MeV\cite{hfb}. Local 
approaches, on the other hand, can be applied to unknown mass regions only with limited confidence. Thus, for example, the neutron rich regions, important for 
various astrophysical processes such as r- or s-process,  are usually studied 
using the microscopic-macroscopic approaches.

Microscopic-macroscopic methods usually use the liquid drop model (LDM) as the 
macroscopic part of the prescription and introduce different corrections
based on microscopic models.
Myers and Swiatecki\cite{MS} included the shell effects by considering the
bunching of the single particle levels. Strutinsky correction\cite{St} takes
care of the shell effects in a fully microscopic way. Finite Range Droplet 
Model (FRDM)\cite{FRDM} is an approach which, besides including Strutinsky 
corrections, pairing and other effects, also modifies the basic liquid drop 
macroscopic part of the mass formula. Another very successful approach  based 
on a microscopic picture  by Duflo and Zuker\cite{DZ} yields an r.m.s. deviation
of 0.373 MeV for  2135 nuclei.

Estimation of binding energy remains a very important task in  nuclear physics.
For example, one of the most important inputs in the study of astrophysical 
reactions is the nuclear ground state binding energy. In nuclei far from the 
stability valley, experimental mass measurements are very scarce and likely to remain so
in  near future. Thus, theoretical prediction of mass remains our 
only guide. In the present work we present a phenomenological formula for 
calculation of binding energy values.

\section{Theory and results}

\subsection{The binding energy formula}

The general approach to a successful nuclear mass formula has been to 
assume the liquid drop model and add shell corrections to it. The liquid drop
model gives a smooth curve while the shell corrections represent the effect
of the single particle states. In the present work, we have employed a 
approach which is close to the Myers and Swiatecki prescription\cite{MS} with 
the shell correction being simulated using a number of parameters.

The original liquid drop model mass formula is known to give incorrect results 
 because of the bunching of the single particle levels. To incorporate this 
effect, we 
have added a term which resembles the form of  the one-body Hamiltonian. 
However, the single particle energies are not constants throughout
the periodic table but
 vary with the neutron and proton numbers of the concerned
nucleus. To simulate this effect, we have used a term of the form
\be B_{bunc}=\sum_i^{1,2}\sum_{\alpha}\epsilon_\alpha^i 
\mathcal{N}^i n^i_\alpha\ee
where $i=1,2$ refer to the neutron or proton, respectively. The number 
of neutrons and protons in the nucleus are given by $\mathcal{N}^1(=N)$ and $\mathcal{N}^2(=Z)$, 
respectively.   We assume a shell model like filling 
of the orbits to obtain the $n^i_\alpha$ values for different nuclei. Thus we 
have
\be  n^i_\alpha=\left\{\begin{array}{lll} 
\mathcal{N}^i_{\alpha+1}-\mathcal{N}^i_{\alpha} & {\rm for} & \mathcal{N}^i> \mathcal{N}^i_{\alpha+1}\\
\mathcal{N}^i-\mathcal{N}^i_{\alpha} & {\rm for} & \mathcal{N}^i_{\alpha}\le \mathcal{N}^i \le \mathcal{N}^i_{\alpha+1}\\
0 & {\rm for} & \mathcal{N}^i< \mathcal{N}^i_{\alpha}\end{array}\right.\nn\ee
The different $\mathcal{N}^i_\alpha$ values are estimated in the following way. 
While fitting the different parameters in the formula for binding energy, we 
have checked the errors at different 
neutron and proton numbers {\em i.e.} for different nuclei. Whenever, the 
errors tend to be large at a particular $N$ or $Z$ value, we assume 
that the single particle structure has changed at that particular number and
new energy level has been occupied. Thus we have assumed $\epsilon_\alpha$ 
values at different neutron or proton numbers ($\mathcal{N}^i_\alpha$) as given  in the 
last five columns of table 1.  This particular form remains an ad hoc 
assumption in our work; yet we find that it can successfully explain the binding energy 
values. 
\begin{table}
\caption{Values of various parameters for the binding energy formula 
obtained in the present calculation. The 
parameters $k_v,k_s,k_W$ are 
dimensionless while $r_0$ is given in $fm$. The rest of the parameters are in 
MeV.}

\begin{center}

\begin{tabular}{cc|ccccc}\hline
     & &$\alpha$          & $\mathcal{N}^1_\alpha$ & $\epsilon^1_\alpha$  & $\mathcal{N}^2_\alpha$ &  $\epsilon^2_\alpha$\\\hline
$a_v$&11.8906 & 1& 8 &0.1284&8&0.2027\\
$a_{surf}$&7.3654& 2& 14&-0.0227&14& 0.0284\\
$k_v$&1.6627& 3& 20& 0.0651&20&0.0889\\
$k_s$&-2.6383&4& 24&0.0306&24&0.0462\\
$W$&35.6813&5&28&0.0394&28&0.0307\\
$k_W$&8.3735&6&32&0.0221&40&0.0191\\
$a_s$&3.1619$\times10^{-8}$&7&40&0.0102&44&0.0256\\
$r_0$&1.0625&8&44&0.0248&50&0.0120\\
$a_c'$&-6.8171$\times10^{-3}$&9&50&0.0089&64&0.0106\\
$a_n$& -1.0645&10&60&0.0138&74&0.0079\\
$a_p$& -0.9508&11&64&0.0093&80&0.0050\\
$a_{np}^2$&3.3594$\times10^{-2}$&12&76&0.0118&82&-0.0001\\
$a_{np}^3$&-3.2543$\times10^{-4}$&13&82&0.0031&88&0.0013\\
$a_\delta$ &4.7686&14&88&0.0079&96&-0.0022\\
$a'_\delta$&-1.4162&15&96&0.0074\\
&&16&122&0.0041\\
&&17&126&0.0043\\
&&18&132&0.0078\\
\hline
\end{tabular}

$\epsilon^1_8$= 0.0197 MeV for $Z\le 34; N\ge 46$\\
$\epsilon^2_2$= 0.0534 MeV  and $\epsilon^2_3$= 0.0312  for $Z\le 22; N\ge 29$\\
\end{center}
\end{table}

Some of the $\mathcal{N}^i_\alpha$ values are easy to understand.  The 
conventional magic numbers and semi-magic numbers do occur in the scheme 
naturally. On the other hand, some other values may be associated
with subshell closures. For example, $\mathcal{N}_\alpha=14$, 32, 64  and 76 
(some in the case of neutrons only) agree with the $1d_{5/2}$, $1p_{3/2}$, 
$2d_{5/2}$ and 
$1h_{11/2}$ subshell closures. Similarly, $N=88$ may be associated with 
a deformed shell closure. Weaker deformed shell closure observed at 
$N=98$ and $Z=100$ are also very close to the value $N_\alpha=96$ that we have 
obtained.
The readers are referred to Refs.\cite{BM1,BM2,MG}
for more details on shell closures and deformed magic numbers. We should also 
mention that Schwierz {\em et al.}\cite{WS} in a study on Woods Saxon 
potential have found indications of smaller gaps at $N,Z=$ 14 and 32 in the 
light mass region. Sorlin  {\em et al.}\cite{Sorlin} have pointed out the 
evidence for subshell closures at 14, 32 and 64.

We have also introduced modifications in a few $\epsilon$ values and the 
$\mathcal{N}^i_\alpha$ values at which they occur for a number of nuclei. This is done 
by again looking at the deviations and checking whether there is any systematic 
pattern. In nuclei with $Z\le 22$ and $N\ge 29$, we find that a better result 
is obtained if we change the $\epsilon^2_2$ and $\epsilon^2_3$  values,
{\em i.e.} for proton  numbers corresponding to 14 and 20, respectively.
Similarly, in nuclei with $Z\le 35$ and $N\ge 46$, we find 
that a better result is obtained if we change the value of the parameter 
$\epsilon^1_8$ (corresponding to neutron number 44)  and assume that the next 
change in single particle level does not correspond to $N=50$.  Interestingly,
this may correspond to the possible disappearance of the magic gap at
$N=50$ in nuclei with $Z\le 35$.
A similar situation appears in nuclei with $Z\le 42$ where the bunching at $N=60$ 
seems to disappear.
The fact that all the above changes occur in neutron rich nuclei is
a signature of rearrangement of single particle levels expected in such regions.

The Binding energy formula used in the present work is given by
\bea
B.E.(N,Z)= B_{LDM}+B_{bunc}+B_{np}+B_{el}\\
B_{LDM}=a_v (1-k_vI^2)A - a_{surf} (1-k_sI^2) A^{2/3} - B_{sym}-B_{Coul}\nn\\
-B_{pair}\eea
\bea
B_{sym}=W|I|(1-k_W|I|)+a_s(N-Z)^5\\
B_{Coul}=\frac{3}{5}\frac{Z(Z-1)e^2}{r_0A^{1/3}}
+a'_c(\frac{Z^2}{A})^{5/2}\\
B_{pair}=a_\delta(\frac{\delta_n}{N^{1/3}}+\frac{\delta_p}{Z^{1/3}})+a_\delta'
\frac{\delta_n\delta_p}{A^{1/3}}\\
B_{np}=a_nN_p+a_pN_n+a_{np}^2(N_p+N_n)^2+a_{np}^3(N_p+N_n)^3\eea

Here, $\delta_{n(p)}$ is 0 or 1, according as $N(Z)$ is even or odd. We write
$I=(N-Z)/A$. The expression for $B_{bunc}$ is already given in eqn (1).
In $B_{np}$, $N_p$ and $N_n$ refers to the number of valence protons and 
neutrons.
The electronic binding energy is estimated by the empirical relation 
\be B_{el}({\rm MeV})=1.44381\times10^{-5}Z^{2.39}+1.55468\times10^{-12}Z^{5.35}\ee

The liquid drop part $B_{LDM}$ contains a number of correction terms along with 
the original BW prescription.
As $B_{bunc}$  contain a large number of $\epsilon^i_\alpha$'s, we have a large 
number of parameters, {\em viz.} fifty in number. If we also consider that the 
nucleon numbers for the bunching of the levels as parameters, the number of 
increases substantially. One obvious 
problem in our approach is  that the nucleon numbers used in the bunching part 
of the formula do not come naturally but have to be introduced from experimental
mass values. Only microscopic calculations can possibly supplement experimental
inputs in this regard.

The bunching term already includes a significant part of the total energy. 
Thus we expect a major modification of the liquid drop model coefficients as 
well as the functional form of some of the  smaller correction terms. We have 
included some different functional forms for the latter terms as described 
below.

The volume and the surface terms as well the contribution of symmetry energy to 
them are retained in the LDM part of the formula. However, the parameters are 
expected to be substantially modified in the present approach. Another major 
modification that we find is in the symmetry energy term $B_{sym}$. The term 
proportional to $|I|$ is the Wigner term that was introduced from the symmetry 
argument. We have introduced two additional terms in the symmetry as is clear 
from the expression. FRDM suggests that the charge asymmetry contribution should
be large in heavier nuclei and uses a term proportional to $(N-Z)$. We find that
a term $(N-Z)^5$ fits the data better in our case.

The Coulomb energy is represented as a sum of two terms. The first one is the 
usual Coulomb term for a charged liquid drop. For the second term, we have 
used a functional of $Z^2/A$. The usual Coulomb contributions include 
an exchange term of the form $Z^{4/3}/A^{1/3}$, proton form factor correction
of the form $Z^2/A$. The FRDM also has contributions from volume and surface
rearrangement terms which depend on $Z$. Lunney \etal\cite{rmp} have also
used a term proportional to $Z^4/A^2$ in the macroscopic part of the energy 
functional related to surface stiffness. We have included a term of the form 
$(Z^2/A)^{5/2}$ as a single term representing the Coulomb corrections. We 
should mention that a nearly comparable fit can be achieved using a form  $(Z^2/A)^2$ also. 

Another addition is the $B_{np}$ term which is a part of the microscopic energy.
It has been demonstrated\cite{Casten,RMF} that 
the number of valence neutrons and protons show a decidedly strong correlation
with n-p interaction and thus with deformation. We have taken a form of the 
type shown in eqn (7). The number of valence particles or holes has been 
calculated from the nearest major shell, i.e. 8, 20, 28, 50 82, 126 and 184.
We have also tried to fit the data using 172 as the last major shell. We find 
that the fit is slightly worse in this case. However, in keeping with the 
expectation that the shell closures are less pronounced in the SHE region,
the binding energy data is not really sufficient to conclude whether the next major closure is
at $N=184$ or 172. On the other hand, as explained later, we have also 
calculated the 
Q-value in $\alpha$-decay where the former value gives decidedly better results
and we have used it in our calculation. The dependence 
of nuclear mass on valence nucleons has already been studied by 
Mendoja-Temis \etal\cite{np} in a different form.

The pairing energy has been shifted so that the value for $B_{pair}$ is zero in
even-even nuclei. We have also assumed a $A^{-1/3}$ dependence for the n-p 
pairing term.

\subsection{Parameters}

The justification of using such a large number of parameters can only be through
a good and simple prediction. One advantage of using this formula is that one 
may use it directly without microscopic correction as the bunching due
to shell effects are included in the $B_{bunc}$ term. We have fitted the 
experimental masses in two steps as explained below.

In the first step we have included
all the experimental mass values in nuclei with $N,Z\ge 8$ from 
the mass table by Audi \etal\cite{mass}. A total of 2140 nuclei have been used 
in 
the fitting procedure yielding an r.m.s. deviation of 0.378 MeV. More importantly, 
there are only 39 nuclei which show deviation larger than 1 MeV and none more 
than 2 MeV. 
In nuclei with mass 
more than 100, only 4 nuclei show a deviation larger than 1 MeV and none more 
than 1.5 MeV. We would like to note that if the modifications in the 
three $\epsilon_i$ values, as explained in the previous subsection, 
are not introduced for neutron rich nuclei, the r.m.s. deviation becomes 0.400 
MeV. In this case we also assume that for the nuclei with $N<22$ and 34, the 
bunching at $N=28$ and 44, respectively, disappears. 

It should be interesting to compare the results for some measurements 
performed later than 2003 and not included in the mass table\cite{mass} 
mentioned above. The masses of  a 
number of nuclei have been measured either for the first time or with increased 
precision in the last few years. We have looked for measurements which are 
completely new or which 
substantially differ from previous values at least in some of the nuclei 
studied. All the measurements from such references have been included for 
completeness. Refs. \cite{new1}$^-$\cite{new16} have
been selected for inclusion. These measurements have been divided in two 
groups. The first group contains those nuclei for which experimental values 
were available in Audi \etal\cite{mass} but later measurements have modified them. There 
are 134 nuclei in this group. In the second group, there are 43 nuclei  whose 
experimental binding energy values have been measured later and were not 
tabulated in Audi \etal\cite{mass}. For both the groups taken together, the r.m.s. error is 
0.438 MeV. For the second group alone, the r.m.s. deviation is 0.517 MeV. 
\begin{figure}[t]
\center
\resizebox{7.4cm}{!}{ \includegraphics{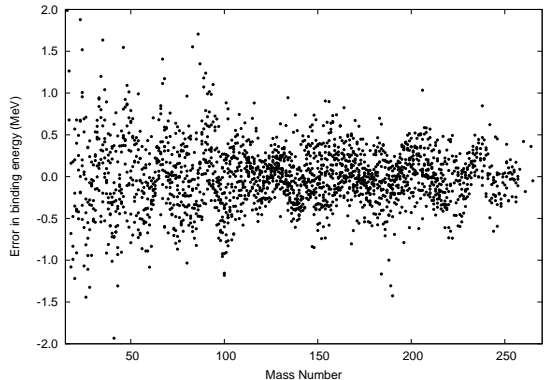}}
\caption{Deviations of binding energy for 2140 nuclei with $N,Z\ge 8$ from the 
predictions of the present formula.}
\end{figure}

\begin{figure}
\center
\resizebox{7.4cm}{!}{ \includegraphics{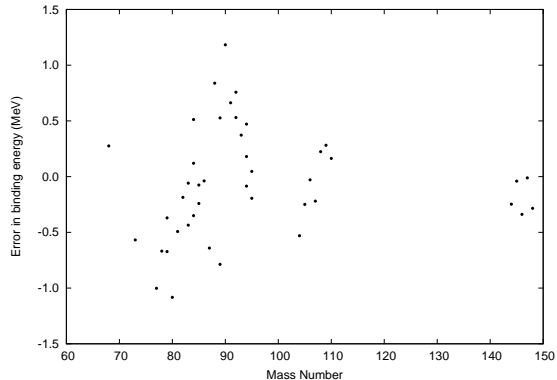}}
\caption{Deviations of binding energy for 43 nuclei whose binding energy
values have recently measured from the predictions of the present formula.}
\end{figure}

In the second step, we have replaced the binding energy values 
of the nuclei belonging to the first group with the improved binding energy 
measurements obtained later wherever available. A fitting procedure for 2140 
nuclei now show 
an r.m.s. deviation of 0.376 MeV. The quality of fit is as good as obtained in 
the first step. The new parameters are very close to the  
fitted values in the first case. The new fitted parameters also describe the 
values in the second group slightly better, the r.m.s. deviation being 0.494 MeV. 
Fig. 1 shows the deviation from the prediction of the (2) formula for the 2140 nuclei while 
Fig. 2 is 
the corresponding figure for the 43 new measurements. The parameters are given in table 1. 
\begin{table}
\caption{Deviation in some global mass formulas}
\begin{center}
\begin{tabular}{lccc}\hline
& No. of nuclei & No. of &  r.m.s. \\
&              & parameters & deviation  \\
&              & & (MeV)     \\\hline
Myers-Swiatecki\cite{adnt}& 1312 & 50 & 0.663 \\
Skyrme-HFB\cite{hfb} & 2149 & 24& 0.581 \\
Duflo-Zuker\cite{DZ} & 2135 & 28&  0.373\\
FRDM\cite{FRDM} & 2135 & 33 & 0.676\\
Present & 2140 & 50 &  0.376\\ \hline
\end{tabular}
\end{center}
\end{table}

In table 2 we compare our results with some other global calculations.
It is clear that our calculation uses more free parameters. Besides, 
a number of parameters in the other approaches are predetermined from 
considerations other than mass measurements. However, as already noted,
the advantage of the present formula is that it is an analytic expression, 
and not dependent on input from any other calculation.

To check the agreement of our results with other predictions, we have chosen 
two isotope chains, $Z=50$ and $Z=110$. The former has been chosen because  
experimental binding energy values for a long chain of isotopes,
from $N=50$ to $N=84$, are available. The other has been chosen to compare the 
predictions in a region where no experimental data is available. The percentage 
deviation values for binding energy are plotted in Fig. 3. The legends AW 
correspond to the experimental or estimated values from the mass 
table\cite{mass}. 
The legends HFB, FRDM and DZ refer to the calculations according to the 
Skyrme Hartree Fock Bogoliubov calculation\cite{hfb}, FRDM 
calculation\cite{FRDM} and Duflo Zuker ten parameter formula\cite{DZ}.
\begin{figure}[t]
\center
\resizebox{!}{!}{ \includegraphics{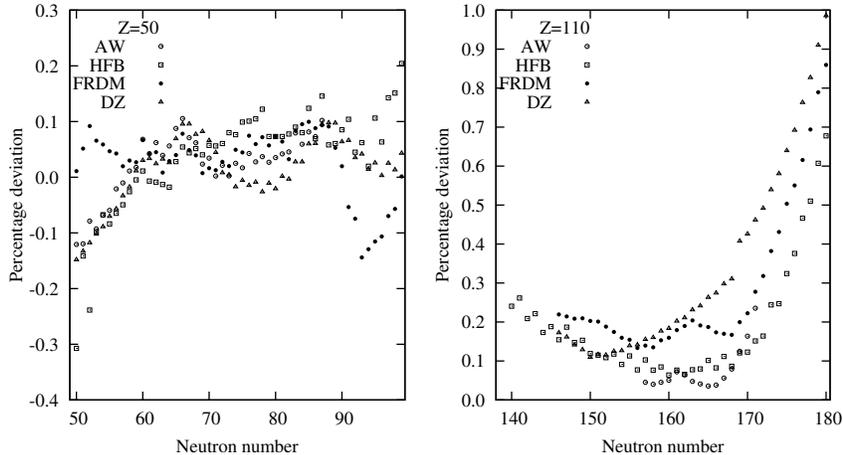}}
\caption{Differences of the present calculation from experimental values and
other prescriptions for $Z=50$ and $Z=110$ nuclei. See text for details.}
\end{figure}

The figure indicates that the present formula agree well for the 
 Z=50 isotope chain. However, the right panel show that there is a systematic  
deviation between the present result and the predictions of HFB, FRDM or 
Duflo-Zuker calculations for $Z=110$ isotopes. The present model
systematically predicts less binding energy for all the nuclei in the chain.
More importantly, beyond $N=172$, the deviations with all the three
theoretical calculations start to rise. When experimental mass values
for these nuclei become available, it will be clear whether the present
prescription needs to be modified. It needs to be remembered that the largest 
 values that have been used in the fitting
procedure are $N=159$ and $Z=108$, as available in Audi \etal\cite{mass}. 

As a further check on the present prescription, we have calculated the alpha 
decay energies in 1091 decays obtained from Audi \etal\cite{mass} augmented by 
the latest results available in the National Nuclear Data Center 
website\cite{web}. The r.m.s. deviation for all these decay energies is 
0.284 MeV, a value consistent with the accuracy in the prediction of binding 
energy in the present work. The deviations in alpha decay energies are plotted
in Fig. 4.
\begin{figure}[t]
\center
\resizebox{7.4cm}{!}{ \includegraphics{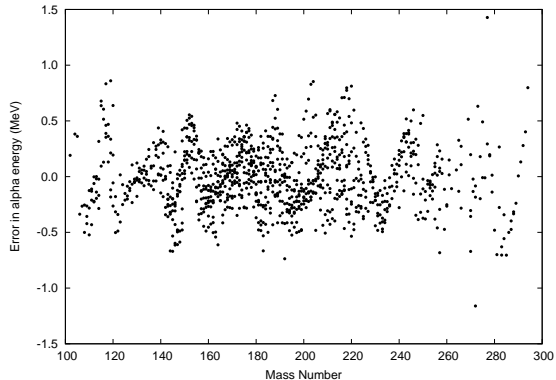}}
\caption{Error in alpha decay energy for 1091 alpha decays}
\end{figure}

It is possible that beyond the highest nucleon numbers with known experimental 
binding energy values, i.e. $N=159$ and $Z=108$, there are other nucleon
numbers where we may have 
to introduce additional bunching. We have looked at the alpha decay energies
for nuclei with larger $N$ or $Z$ values and have tried to fit with
different $\epsilon$ parameters at different $N$ or $Z$ numbers.  
A least square fitting did not yield a substantial improvement in the 
alpha decay energies. However, as the number of observed decays in this
region is not large, it is not possible to conclude positively about
possible change in bunching beyond the above $N$ or $Z$ values.
We have already noted that alpha decay Q-values indicate that  $N=184$ is a 
better closed core than 172.

\subsection{Correspondence with the liquid drop model}

We see that the usual liquid drop model parameters are modified to a large 
extent in the present calculation. 
It is difficult to compare the standard values of parameters with the specific 
form in (1). Parts of all the conventional liquid drop model terms, such 
as the volume, surface, symmetry and Coulomb energies, are included in the 
bunching part of the formula. Thus the values of the different parameters in 
the present work differ substantially from the values obtained from their 
counterparts in conventional calculations. Even a parameter like $r_0$, which 
seems to take a value in conformity with other results, is actually very much 
dependent on the form of the functional chosen for the second term in 
$B_{Coul}$.  

It is possible to visualize our results in the following way. In $N=Z$ nuclei,
there is no contribution from symmetry energy. We would like to interpret the
difference in contribution from the proton and the neutrons to $B_{bunc}$ 
in such nuclei as essentially a part of the Coulomb energy in the
conventional LDM. Thus we have written the Coulomb energy in $N=Z$ nuclei,
neglecting the difference between the neutron and proton energies in $B_{np}$,
as $E_{Coul}=B_{Coul}+B_{bunc}^p-B_{bunc}^n$. This quantity we have calculated 
for $N=Z$ nuclei up to 108 from the proposed binding energy formula 
using the parameters deduced above and have plotted 
it with a simple conventional form consisting of a direct term and a 
diffuseness correction, {\em i.e.}
\be \frac{3}{5}\frac{e^2Z^2}{r_0A^{1/3}}+f_p\frac{Z^2}{A}\ee
Royer and Rousseau\cite{Royer1} 
pointed out that the Coulomb exchange term is unnecessary if the diffuseness 
correction is included. 
Fitting $E_{Coul}$ with the above expression, we obtain $r_0$=1.2114 fm and 
$f_p=1.0395$ MeV which are
within the range of usual values\cite{Royer}. 
The Coulomb energy $E_{Coul}$ is better described using a functional of the form
\be \frac{3}{5}\frac{e^2Z^2}{r_0A^{1/3}}+f_p\frac{Z^2}{A}
+a_{ch}\frac{Z^4}{A^2}\ee

In even-even $N=Z$ nuclei, it is possible to extract the volume and surface 
term coefficients from the mass formula. The sum of volume and surface terms
is obtained by adding $E_{Coul}$ to the total B.E. from eqn (2) taking the 
microscopic corrections into account and fitted with a functional of the form 
\be \alpha_vA-\alpha_sA^{2/3}\ee
The values of the coefficients
$\alpha_v$ and $\alpha_s$ have been obtained in this way. We have considered 
only nuclei for which the  microscopic correction terms from \cite{FRDM} are 
available. 

With the equivalent volume, surface and Coulomb energies given by the 
prescriptions (10) and (11), it is possible to estimate the symmetry energy
contained in the values obtained from eqn (2). 
We have calculated the symmetry energy for all the even even nuclei among the
2140 which we have fitted originally taking into account the microscopic 
corrections
from M\"{o}ller \etal\cite{FRDM}. 
There are various possible expressions for the
symmetry energy in LDM and we have taken a functional of the form
\be -a_vk_vI^2A+a_sk_sI^2A^{2/3}-W|I|+c_s(N-Z)\ee
the last term being used in the FRDM. The different parameters have been 
estimated by least square fitting. 

The interpretation of pairing energy terms for odd-Z or odd-N terms are
straightforward. The coefficient of the $1/\sqrt{N(Z)}$ term is
4.77 MeV, a value very close to the accepted value of 4.8 MeV. The 
n-p pairing term is usually taken in a different form from the present 
work and cannot be directly compared. Its effect is of course small.

Following the procedure described above, we get the following expression for 
binding energy in terms of the deduced parameters:
\bea
B({\rm MeV})=&15.2893(1-1.8735I^2)A-16.4662(1-2.4595I^2)A^{2/3}\nn\\
&-\frac{3}{5}\frac{e^2Z^2}{1.3662A^{1/3}}+0.3639\frac{Z^2}{A}-0.0421\frac{Z^4}{A^2}\nn
\\
&-28.0438|I|-0.4857(N-Z)-B_{pair}-E_{shell}\eea
where $E_{shell}$ values are the microscopic corrections from M\"{o}ller 
\etal\cite{FRDM}.
It is seen that the coefficients in the above formula are within the
accepted ranges for their counterparts in the liquid drop model. We note
that the parameters in the above expression have been obtained fitting, not 
the experimental values, but the values predicted by eqn (2).
Using the above formula, we find that the r.m.s. deviation from experimental 
values is obtained as 0.969 MeV for the nuclei in Audi \etal\cite{mass} Thus, we may 
conclude that the present prescription is equivalent to a conventional
LDM formula which takes shell corrections into account.

\section{Summary}

A new phenomenological formula has been proposed for ground state binding 
energies of nuclei. Two additional components have been included in the proposed
formula in addition to the conventional liquid drop model. The contribution of 
the bunching of single particle levels has been included through a term 
resembling the one-body Hamiltonian. No additional shell correction from 
any external calculation has 
been considered. The effect of n-p interaction and deformation has been included
through a function of the number of valence nucleons. A total of 50 parameters has been used 
in the present calculation. The r.m.s. deviation in the ground state binding 
energy  for 2140 nuclei comes out to be 0.376 MeV. As an additional check on the formula, decay energies for 1091 alpha decaying nuclei has been calculated 
which yields an r.m.s. deviation of 0.284 MeV. As the bunching term incorporates
a large part of the conventional liquid drop model terms, the parameters in the 
present work differ substantially form their standard values. The present 
formula may be reduced to the form of a conventional liquid drop model with
parameters close to standard values. 

\section*{Acknowledgment}

This work is carried out with financial assistance of the UGC sponsored
DRS Programme of the Department of Physics of the University of Calcutta.
The author would like to thank T.K. Das for reading the manuscript and 
making helpful suggestions.

\end{document}